\newcommand{\be}{\begin{equation}}
\newcommand{\ee}{\end{equation}}
\newcommand{\bea}{\begin{eqnarray}}
\newcommand{\eea}{\end{eqnarray}}
\newcommand{\ba}[1]{\begin{array}{#1}}
\newcommand{\ea}{\end{array}}
\documentclass[aps,twocolumn,showpacs,floatfix]{revtex4}
\usepackage{amsmath}
\usepackage{amsfonts}
\usepackage{amssymb}
\usepackage{graphicx}
\usepackage{bbm}
\begin{document}
\title{Quantum Entanglement in Coupled Lossy Wave\-guides}
\author{ Amit Rai, Sumanta Das and G. S. Agarwal }
\affiliation{Department of Physics, Oklahoma State University,
Stillwater, Oklahoma 74078, USA}
\date{\today}
\begin{abstract}

We investigate the viability of coupled wave\-guides as basic units
of quantum circuits. In particular, we study the dynamics of
entanglement for the single photon state, and single mode squeezed
vacuum state. We further consider the case of entangled inputs in
terms of the two mode mode squeezed vacuum states and the two photon
NOON state. We present explicit analytical results for the measure
of entanglement in terms of the logarithmic negativity. We also
address the effect of loss on entanglement dynamics of waveguide
modes. Our results indicate that the waveguide structures are
reasonably robust against the effect of loss and thus quite
appropriate for quantum architectures as well as for the study of
coherent phenomena like random walks. Our analysis is based on
realistic structures used currently.

\end{abstract}

\pacs{42.82.Et, 42.50.-p, 42.50.Ex}

\maketitle

\section{Introduction}

Discrete optical systems like coupled wave\-guides are known to be
extremely efficient in manipulating the flow of light and have been
investigated extensively in the last two decades
\cite{Peschel,Eisenberg,Pertsch,Dimtri}. Many key quantum effects
like quantum interference, entanglement and quantum walk has been
investigated in these systems \cite{Longhi3,Longhi4,
hagai,Zee,rai1}. For example, using coherent beam Peretes et al.
\cite{hagai} have observed quantum walk effects in a system
consisting of large number of wave\-guides. In another experiment,
Bromberg et al. investigated the quantum correlations in GaAs
waveguide arrays \cite{Zee} using two-photon input states. In
particular, they considered both the separable and entangled two
photon state and observed various features associated with quantum
interference. In addition, the coupled wave\-guides arrays have been
used to study the discrete analogue of the Talbot effect
\cite{iwanow}. Note that, entanglement between the wave\-guide modes
and behavior of nonclassical light in coupled wave\-guides have also
attracted a great deal of interest \cite{Longhi1,Longhi2,rai2}. In a
recent experiment Politi et. al. \cite{politi} have shown how a CNOT
gate can be implemented on a single Silicon chip using coupled
silica wave\-guides, thus showing possible application of
wave\-guides in quantum computation. They also observed two photon
interference in these coupled wave\-guides. In a following
experiment \cite{Matthews, Berry} coupled silica wave\-guides was
used to generate a multi\-mode interferometer on an integrated chip.
It was further shown that these interferometers can be used to
generate arbitrary quantum circuits. They also showed that two and
four photon entangled states similar to NOON states \cite{dow} can
be generated on the silicon chip. All these studies have hence given
a new impetus in the field of quantum information processing and
quantum optics with wave\-guides. Note that the entanglement between
wave\-guide modes is at the heart of many of these experiments. In
particular, for effective use of these wave\-guide circuits in
quantum computation and communication tasks sustainability of
generated entanglement is very important \cite{niel,bennette}. In
light of this, it is imperative to study entanglement in
wave\-guides using quantitative measures for entanglement. This is
the main purpose of the present study. Moreover, in practice the
wave\-guides are not completely lossless. Thus an immediate question
of interest would be how does this loss affects the entanglement in
the waveguide modes ? It is well known that entanglement is quite
susceptible to decoherence \cite{zurek} and thus the above question
bears immense interest in context to quantum information processing
using wave\-guides. Further it is important to understand the role
of loss in coherent phenomena like quantum random walk \cite{hagai,pathak}.\\
\indent In this paper we investigate these in a simple system of two
single mode wave\-guides, which are coupled through the overlap of
evanescent fields. This simple system serves as a unit or the basic
element for constructing a quantum circuit \cite{Alberto}. The input
light to the coupled waveguide system is usually produced by a
parametric down-conversion process at high and low gain which
produces important nonclassical states of light like the squeezed
and the single photon states respectively. Thus the input is quite
naturally a squeezed state specially at high gain. Behavior of
photon number states such as the single photon state and the NOON
state have also been investigated in these systems
\cite{politi,Matthews,Alberto}. We thus consider a variety of
nonclassical input states like squeezed states and photon number
states which have been extensively investigated in couple
wave\-guide system and study their respective entanglement dynamics.
We quantify the evolution of entanglement in terms of logarithmic
negativity and present explicit analytical results for both squeezed
and number state inputs. We further investigate the question of
possible effects of loss on the entanglement dynamics in
wave\-guides by considering lossy waveguide modes. We find that in
this case, for both number state inputs as well as squeezed state
inputs, entanglement shows considerable robustness against
loss. \\
\indent{} The organization of the paper is as follows: In Sec. II,
we describe the model and derive analytical result for the field
modes of the coupled wave\-guide system. In Sec. III, we study the
evolution of entanglement for two classes of photon number states,
(A) separable single photon state $|1,1\rangle$ and (B) entangled
two-photon NOON state. We quantify the degree of entanglement of
these states by using the logarithmic negativity. In sec. IV , we
then study the time evolution of entanglement by evaluating the
logarithmic negativity for two classes of squeezed input states (A)
separable two mode squeezed state and (B) entangled two mode
squeezed state. The effect of loss in wave\-guides on the
entanglement dynamics is then discussed in Sec. V. Finally we
summarize our results in section VI with a future outlook.

\section{ The Model }

We consider a system with two single mode wave\-guides, coupled
through nearest-neighbor interaction as shown in Fig. 1. Let $a$ and
$b$ be the field operators for the modes in each wave\-guide. These
obey bosonic commutation relations $[a,a^\dagger]=1$; $(a
\rightarrow b)$. The Hamiltonian describing the evanescent coupling
between the wave\-guide mode in such a system of two coupled
wave\-guides can be derived using the coupled mode theory
\cite{saleh,Marcuse}. The coupling among the wave\-guides is
incorporated in this framework by treating it as a perturbation to
the mode amplitudes. It is assumed that the presence of the second
wave\-guide perturbs the medium outside the first wave\-guide. This
creates a source of polarization outside the first wave\-guide,
which thereby leads to modification of the amplitude of the mode in
it. Further, the amplitude of the modes in each wave\-guide is
assumed to be a slowly varying function of the propagation distance.
Moreover, in this perturbative approach the coupling does not effect
the propagation constant or transverse spatial distribution of the
wave\-guide modes. The field of the first wave\-guide has a similar
effect on the second wave\-guide. Under these assumptions, the field
mode of the composite structure are governed by the Helmholtz
equation which gives two coupled first order differential equations
which can be solved to obtain the time evolution of field modes in
the coupled wave\-guide structure. The corresponding description for
the nonclassical light can be studied by quantizing the field
amplitudes as has been done in the work of Lai et. al. \cite{buzek}.
Following an approach similar to that developed by Lai et. al., we
can write the corresponding quantum mechanical Hamiltonian for the
coupled waveguide as

\begin{eqnarray}
\label{1}
 H = \hbar\omega (a^\dagger a +b^\dagger  b )+
  \hbar  J(a^\dagger
 b + b^\dagger
 a )~,
\end{eqnarray}
\begin{figure}[htp]
\scalebox{0.38}{\includegraphics{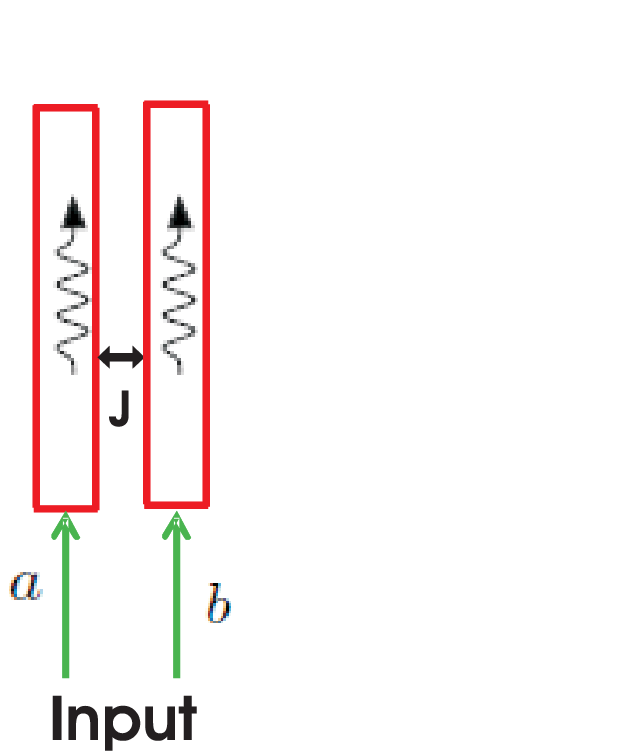}} \caption{(Color online)
\label{Fig2(a)}Schematic diagram of a coupled wave\-guide system.
The parameter
 $J$ gives the coupling between the waveguide modes and $\gamma$ is the loss rate. }
 \end{figure}

\noindent where the first two terms correspond to the free energy of
the wave\-guide modes and  the last two terms account for the
evanescent coupling between the  wave\-guide modes with $J$ as the
coupling strength. The coupling $J$ depend on the distance between
the wave\-guides. The input to the coupled wave\-guide system can be
in a separable or an entangled state. Let $\gamma$ be the loss rates
of the modes $a$ and $b$. The loss $\gamma$ arises from the loss in
the material of the wave\-guide. Table I below gives the
experimental values of coupling parameter $J$ and loss $\gamma$ for
different wave\-guide systems.

\begin{widetext}

\begin{center}
    \begin{tabular}{ | l | l | l | l |}
    \hline
    Wave\-guide Type & Coupling parameter
   $J$ ($sec^{-1}$) & Loss $\gamma$ ($sec^{-1}$) & $\gamma$/$J$\\ \hline
    Lithium Niobate ($LiNbO_{3}$)
    & $1.83\times 10^{10}$ - $4.92\times 10^{10}$
 & $3 \times
10^9$ & $1/7$-$1/20$ \\ \hline
    AlGaAs  & $2.46 \times 10^{11}$  & $2.7 \times 10^{10}$ & $1/10$  \\
    \hline
    Silica & $1.53 \times 10^{11}$ & $3 \times 10^9$ & $1/50$ \\
    \hline

      \end{tabular}

      \begin{table}[h!b!p!]
\caption{Approximate values of some of the parameters used in
wave\-guide structures \cite{Stegeman,Aitchison,Kutter}. The loss,
usually quoted in $dB/cm$, for different wave\-guides is converted
to frequency units used in this paper by using the formula, $10
\hspace{0.1 cm} Log (\frac{P_{out}}{P_{in}}) \equiv 10 \hspace{0.1
cm} Log (e^{-2 \gamma/c} )$, where $P_{in}$ is the input power,
$P_{out}$ is the power after traveling unit length.}

\end{table}

\end{center}

\end{widetext}

\noindent As known the silica wave\-guides have very little
intrinsic loss and should be preferable in many applications.
Nevertheless the loss is to be included as this could be detrimental
in long propagation for example in the study of quantum random
walks. Since the two wave\-guides are identical, we have taken the
loss rate of both the modes to be the same. We can model the loss in
wave\-guides in the framework of system-reservoir interaction well
known in quantum optics and is given by, \\
\begin{eqnarray}
\label{2} \mathcal{L}\rho & =
&-\frac{\gamma}{2}(\hat{a}^{\dagger}\hat{a}\rho-2\hat{a}
\rho\hat{a}^{\dagger}+\rho\hat{a}^{\dagger}\hat{a})\nonumber\\
& & -\frac{\gamma}{2}(\hat{b}^{\dagger}\hat{b}\rho-2\hat{b}
\rho\hat{b}^{\dagger}+\rho\hat{b}^{\dagger}\hat{b})~,
\end{eqnarray}

where $\rho$ is the density operator corresponding to the system
consisting of fields in the modes a and b. The dynamical evolution
of any measurable $\langle O \rangle $ in the coupled wave\-guide
system is then governed by the quantum-Louiville equation of motion
given by,
\begin{equation}
\label{3} \dot{\rho} = -\frac{i}{\hbar}[H, \rho]+\mathcal{L}\rho
\end{equation}

where $\langle \dot{O} \rangle = {\rm Tr}\{O\dot{\rho}\}$, the
commutator gives the unitary time evolution of the system under the
influence of coupling and the last term account for the loss. Note
that in absence of loss (loss\-less wave\-guides) the time evolution
of the field operators can be evaluated using the Heisenberg
equation of motion and is given by, \bea \label{4}
a(t) & = & a(0)\cos(Jt)-ib(0)\sin(Jt)\nonumber\\
b(t) & = & b(0)\cos(Jt)-ia(0)\sin(Jt). \eea Next we will study the
entanglement characteristics of photon number and squeezed input
states as they propagate through the wave\-guides. To keep the
analysis simple in the next few sections we consider the case of
loss\-less wave\-guide modes ($\gamma = 0$). We defer the discussion
of loss on entanglement to Sec V.

\section{Evolution of entanglement for non-Gaussian input states}
\noindent In this section we study the dynamics of entanglement for
photon number input state. We quantify the entanglement of the
system by studying the time evolution for the logarithmic negativity
\cite{Vidal, Plenio1, Plenio2}. For a bipartite system described by
the density matrix $\rho$ the logarithmic negativity is
\begin{eqnarray}
&& E_\mathcal{N}(t)  =  \log_{2}
\parallel\rho^{T} \parallel ,\nonumber\\
&& \parallel\rho^{T}\parallel = (2N(\rho)+1)\label{eq4}~,
\end{eqnarray}
\noindent where $\rho^{T}$ is the partial transpose of $\rho$ and
the symbol $\parallel \hspace{0.2 mm} \parallel$ denotes the trace
norm. Also $N(\rho)$ is the absolute value of the sum of all the
negative eigenvalues of the partial transpose of $ \rho$. The log
negativity is a non-negative quantity and a non-zero value of
$E_\mathcal{N}$ would mean that the state is entangled.

\subsection{Separable photon number state as an input}

\noindent We first consider the case when there is no loss and hence
we set $\gamma = 0$. We assume that the input is in a separable
state. Further, for studying the entanglement dynamics for photon
number states we first consider the case of a single photon input in
each wave\-guide. Thus the initial state is

\begin{eqnarray}
 |\psi (0)\rangle = |1, 1\rangle.\label{004}
\end{eqnarray}
\noindent Using Eq.~(\ref{4}) we can show that a single photon input
state given by $ |\psi (0)\rangle$ evolves into a state~:
\begin{eqnarray}
|\psi (t) \rangle \rightarrow \alpha_1 |2,0\rangle+\beta_1
|1,1\rangle + \delta_1 |0,2\rangle~.\label{eq2a}
\end{eqnarray}
\noindent The coefficients $\alpha_1 $, $\beta_1$ and $\delta_1 $
are given by :
\begin{eqnarray}
&& \alpha_1 \equiv -i  \sin( 2 J t)/\sqrt{2}  ,\nonumber\\
&& \beta_1 \equiv  \cos( 2 J t),\nonumber\\
&& \delta_1 \equiv  -i  \sin( 2 J t)/\sqrt{2} ~.\label{eq3}
\end{eqnarray}
\noindent The density matrix corresponding to the state in
(\ref{eq2a}) can be written as :
\begin{eqnarray}
\rho  &  = & |\psi (t) \rangle  \langle \psi (t)|\nonumber\\ &=&
|\alpha_1|^2 |2,0\rangle \langle 2,0| +|\beta_1|^2 |1,1\rangle
\langle
1,1| + |\delta_1|^2 |0,2\rangle \langle 0,2|\nonumber\\
& & + \alpha_1 \beta_1^* |2,0\rangle \langle 1,1| + \delta_1
\beta_1^* |0,2\rangle \langle 1,1|+
\alpha_1 \delta_1^*  |2,0\rangle \langle 0,2| \nonumber\\
& & + \beta_1 \alpha_1^* |1,1\rangle \langle 2,0| + \beta_1
\delta_1^* |1,1\rangle \langle 0,2|+
 \delta_1 \alpha_1^*  |0,2\rangle \langle 2,0|~.\nonumber\\
\label{eq10}
\end{eqnarray}

\noindent Using Eq.~(\ref{eq4}) and the above equation, we can show
that the log negativity $E_\mathcal{N}$ is given by:

\begin{eqnarray}
E_\mathcal{N} &&  = \log_{2}(1+2N(\rho)) ~,\nonumber\\
&& =\log_{2}(1+2|(\alpha_1 \beta_1+ \alpha_1 \delta_1+\delta_1
\beta_1)|)\label{eq6}~.
\end{eqnarray}
\begin{figure}[htp]
 \scalebox{0.48}{\includegraphics{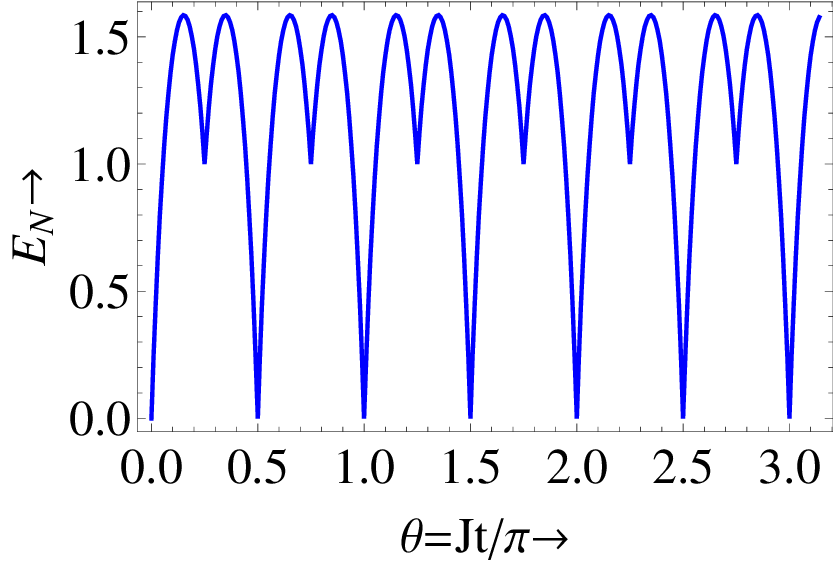}}
 \caption{\label{Fig1(a)}Time evolution of log negatively for a single photon input state (\ref{004}).}
\end{figure}

\noindent In Fig.~(\ref{Fig1(a)}) we show the time evolution of
$E_\mathcal{N}$ for the single photon input state $|1,1\rangle$. We
would like to emphasize that the values of $\theta $ studied here
are very similar to the ones employed in the recent experiments
\cite{Zee, politi}. At time $t=0$, we begin with a separable input
state and thus the value of log negativity is $E_\mathcal{N}=0$. The
entanglement quantified by the log negativity increases with time
and attains a maximum value of $1.58$ for $ \theta \simeq 0.15 $. In
this case the single photon state evolves into a maximally entangled
state given by: $|\psi_m \rangle \equiv e^{{-i \pi/2}}( |2,0\rangle+
|0,2\rangle) +|1,1\rangle/\sqrt{3} $. Further, for $ \theta = 1/4 $,
we get an analog of the well known Hong\textendash Ou\textendash
Mandel interference \cite{hong}. Note that in this case the
logarithmic negativity $E_{\mathcal{N}} $ attains a value of $1$
which is less than the corresponding value of $E_{\mathcal{N}} $ for
the maximally entangled state $|\psi_m \rangle$. In addition, for $
\theta = 1/2 $, we find that $E_{\mathcal{N}}$ vanishes and the
state at this point is $e^{i \pi} |1,1\rangle$. At later times, we
see a periodic behavior which can be attributed to the
inter-wave\-guide coupling $J$. \noindent We next consider the case
where we have two photons in one wave\-guide and none in the other
input. Thus the initial state can be written as :

\begin{eqnarray}
 |\varphi (0)\rangle = |2, 0\rangle~.\label{eq2zz}
\end{eqnarray}
\noindent Again using Eq.~(\ref{4}) we find that the $|\varphi
(0)\rangle$ evolves into a state
 given by~:
\begin{eqnarray}
|\varphi (t) \rangle \rightarrow \alpha_2 |2,0\rangle+\beta_2
|1,1\rangle + \delta_2 |0,2\rangle~.\label{eq2z}
\end{eqnarray}
\noindent The coefficients $\alpha_2 $, $\beta_2$ and $\delta_2 $
are given by :
\begin{eqnarray}
&& \alpha_2 \equiv \cos( J t)^2 ,\nonumber\\
&& \beta_2 \equiv -\sqrt{2} i \cos( J t) \sin( J t),\nonumber\\
&& \delta_2 \equiv  -\sin( J t)^2~.\label{eq3z}
\end{eqnarray}

\begin{figure}[htp]
 \scalebox{0.48}{\includegraphics{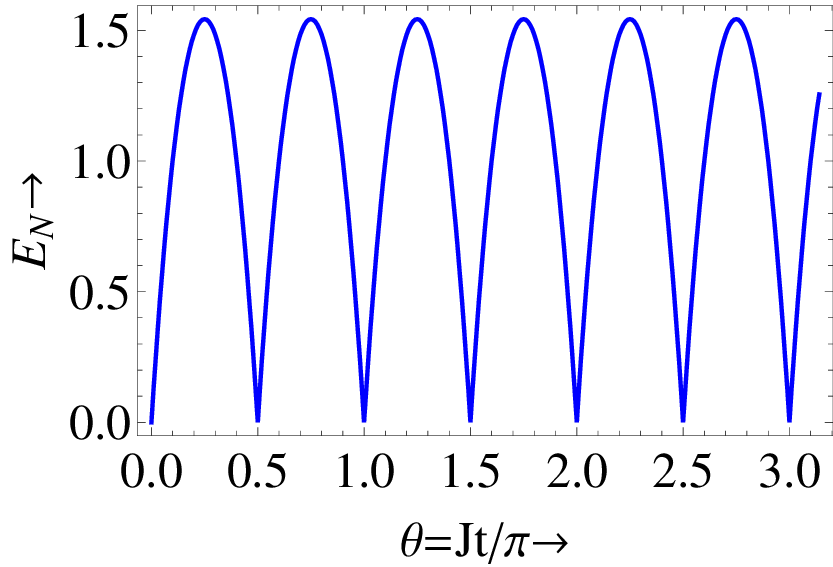}}
 \caption{{\label{eq2}} The behavior of log negatively for the
state (\ref{eq2zz}) as function of $ \theta= Jt/\pi$. }
 \end{figure}

\noindent Using a similar procedure as discussed above we can
evaluate the log negativity $E_\mathcal{N}$ for the state in
Eq.~(\ref{eq2z}). We show the result for the log negativity in
Fig.~(\ref{eq2}). In this case we find that the log negativity
increases and attains a maximum value of $1.54$. After reaching the
maximum value the log negativity decreases and eventually becomes
equal to zero. Thus the state becomes disentangled at this point of
time. At later times we see a periodic behavior and the system gets
entangled and disentangled periodically. Clearly the entanglement
dynamics of the states (\ref{004}) and (\ref{eq2zz}) are different.
Unlike the earlier case for the $|1,1\rangle$ input state, we don't
see any interference effects in this case \cite{Zee}.

\subsection{Entangled photon number state as an input}

\noindent Next we consider the entangled state prepared in a two
photon NOON state \cite{dow} as our initial state :
\begin{eqnarray}
|\phi (0) \rangle  & = & \frac{(|2,0\rangle +
|0,2\rangle)}{\sqrt{2}}~.\label{eq5}
\end{eqnarray}
\begin{figure}[htp]
 \scalebox{0.58}{\includegraphics{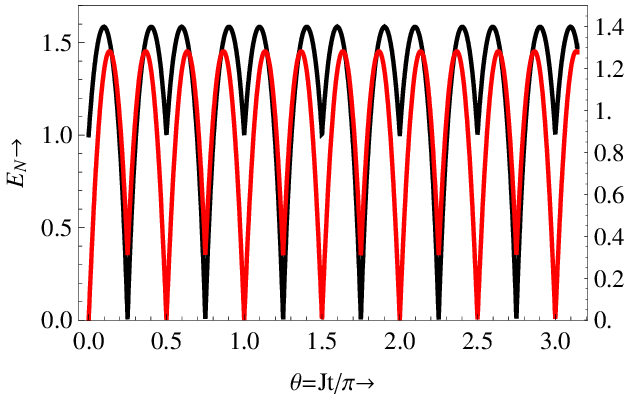}}
 \caption{\label{Fig2(a)} The behavior of log negatively as function of $ \theta= Jt/\pi$. The black
  curve shows the result for two photon NOON state while the red curve shows the result($E_{N}-1$) for the four photon NOON state}
 \end{figure}

\noindent As shown in the black curve of Fig.~(\ref{Fig2(a)}), the
value of $E_{\mathcal{N}} $ at time $t=0$ is equal to $1$ which
indicates entanglement. The log negativity $E_{\mathcal{N}}$ in
Fig.~(\ref{Fig2(a)}) shows a behavior that is similar to the result
for the $|1,1\rangle$ state shown in  Fig.~(\ref{Fig1(a)}). Also
note the shift of $\pi/4$ between the results in
Figs.~(\ref{Fig1(a)}) and (\ref{Fig2(a)}). As in the case of single
photon input state $|1,1\rangle$, the initial state evolves into a
maximally entangled state corresponding to a value of
$E_{\mathcal{N}}$ which is equal to $1.58$. In addition, for $
\theta = 1/2 $, we again see a signature of quantum interference
such that the probability of getting the single photons at each of
the output port vanishes \cite{Zee}. The logarithmic negativity
$E_{\mathcal{N}} $ at this point is equal to $1$. At later times the
entanglement shows an oscillatory behavior and the system gets
periodically entangled and disentangled. For an arbitrary NOON state
: $|\psi_ {in}\rangle  = ( |N,0\rangle+ |0,N\rangle)/\sqrt{2}$ the
state at time $t$ is $|\psi_ {out}\rangle  = (\sum \beta_{k}
|k,N-k\rangle)$ where $ \beta_{k}$ can be written in terms of the
binomial coefficient (see Eq.~(\ref{app})). For completeness, we
give the details for the calculation of time evolution of $|\psi_
{in}\rangle $ in Appendix A. The density matrix corresponding to the
state $|\psi_ {out}\rangle$ can be written as : $\rho_ {out} = \sum
\beta_{k} \beta_{m}^{*} |k,N-k\rangle \langle m,N-m|$. Taking the
partial transpose of $\rho_ {out}$, we get $\rho_ {out}^{T} = \sum
\beta_{k} \beta_{m}^{*} |k,N-m\rangle \langle m,N-k|$. Further it
can be proved that ${(\rho_ {out}^{T})}^2$ is a diagonal matrix and
the eigenvalues of ${(\rho_ {out}^{T})}^2$ is of the form:
$|\beta_{k}|^2 |\beta_m|^2$. Thus the negative eigenvalues of
${\rho_ {out}^{T}}$ are of the form $|\beta_{k}| |\beta_m|$ $(k\neq
m)$ and the log negativity $E_\mathcal{N}$ can be written as:

\begin{eqnarray*}
E_\mathcal{N} &&  = \log_{2}(1+2N(\rho)) ~,\nonumber\\
&& =\log_{2}(1+2\sum_{k\neq m} |\beta_k| |\beta_m|)~.
\end{eqnarray*}

\noindent We can use the above equation to study entanglement
dynamics for the $N$ photon NOON state. The red curve in Fig. 4
shows the result for the four photon NOON state. As earlier, the
value of $E_{\mathcal{N}} $ at time $t=0$ is equal to $1$ which
indicates entanglement. The curve for four photon NOON state also
shows quantum interference effect. Further, the logarithmic
negativity never becomes zero in this case and hence the initially
entangled state remains entangled for later times.

\section{Evolution of entanglement for Gaussian input states}

\subsection{Separable two mode squeezed state as an input}

We next study the generation and evolution of entanglement for the
case of squeezed input states. For this purpose we first consider a
separable squeezed input state coupled to the modes $a$ and $b$ of
the waveguide given by, \be \label{20} |\zeta\rangle =
|\zeta_{a}\rangle\otimes |\zeta_{b}\rangle ; \ee where
$|\zeta_{a}\rangle(|\zeta_{b}\rangle)$ are single mode squeezed
states defined as, \be \label{21} |\zeta_{a}\rangle =
\exp(\frac{r}{2}\{a^{\dag 2}-a^{2}\})|0\rangle;\quad (a \rightarrow
b). \ee where $r$ is taken to be real. It is well known that a two
mode squeezed state like $|\zeta\rangle$ can be completely
characterized by its first and second statistical moments given by
the first moment : $(\langle x_{1}\rangle, \langle p_{1}\rangle,
\langle x_{2}\rangle, \langle p_{2}\rangle)$ and the covariance
matrix $\sigma$. The squeezed vacuum state falls under the class of
Gaussian states. It is to be noted that evolution of Gaussian states
has been studied for many different model Hamiltonians
\cite{Halliwell,vitali11,Plenio111,Agarawal112}. We focus on the
practical case of propagation of light produced by a down converter
in coupled wave\-guides which currently are used in quantum
architectures and quantum random walks. Note that since the first
statistical moments can be arbitrarily adjusted by local unitary
operations, it does not affect any property related to entanglement
or mixedness and thus the behavior of the covariance matrix $\sigma$
is all important for the study of entanglement. The measure of
entanglement for a Gaussian state is best characterized by the
logarithmic negativity $E_{\mathcal{N}}$, a quantity evaluated in
terms of the symplectic eigenvalues of the covariance matrix
$\sigma$ \cite{duan, simon}. The elements of the covariance matrix
$\sigma$ are given in terms of conjugate observables, $x$ and $p$ in
the form, \bea \label{23} \sigma =
\left[\begin{array}{ccc} \alpha & &\mu \ \\
\\
\mu^{T} & &\beta\end{array}\right]; \eea where $\alpha, \beta$ and
$\mu$ are $2\times2$ matrices given by, \be \label{24} \alpha =
\left[\begin{array}{cc} \langle x^{2}_{1}\rangle & \langle \frac{x_{1}p_{1}+p_{1}x_{1}}{2}\rangle \\
\\
\langle \frac{x_{1}p_{1}+p_{1}x_{1}}{2}\rangle & \langle
p^{2}_{1}\rangle\end{array}\right]; \ee \be \label{25} \beta =
\left[\begin{array}{cc} \langle x^{2}_{2}\rangle & \langle \frac{x_{2}p_{2}+p_{2}x_{2}}{2}\rangle \\
\\
\langle \frac{x_{2}p_{2}+p_{2}x_{2}}{2}\rangle & \langle
p^{2}_{2}\rangle\end{array}\right]; \ee \be \label{26} \mu =
\left[\begin{array}{cc} \langle\frac{x_{1}x_{2}+x_{2}x_{1}}{2}\rangle & \langle \frac{x_{1}p_{2}+p_{2}x_{1}}{2}\rangle \\
\\
\langle \frac{x_{2}p_{1}+p_{1}x_{2}}{2}\rangle &
\langle\frac{p_{1}p_{2}+p_{2}p_{1}}{2}\rangle
\end{array}\right].\ee

Here $x_{1},x_{2}$ and $p_{1},p_{2}$ are given in terms of the
normalized bosonic annihilation (creation) operators $a
(a^{\dagger})$, $b (b^{\dagger})$ associated with the modes $a$ and
$b$ respectively, \bea \label{21b}
x_{1} &=& \frac{(a+a^{\dag})}{\sqrt{2}},\quad x_{2} = \frac{(b+b^{\dag})}{\sqrt{2}}; \nonumber\\
p_{1} &=& \frac{(a-a^{\dag})}{\sqrt{2}i}, \quad p_{2}
=\frac{(b-b^{\dag})}{\sqrt{2}i} \eea The observables, $x_{j},p_{j}$
satisfy the cannonical commutation relation $[x_{k},p_{j}] =
i\delta_{kj}$. The condition for entanglement of a Gaussian state
like $|\zeta\rangle$ is derived from the PPT criterion \cite{simon},
according to which  the smallest symplectic eigenvalue
$\tilde{\nu}_{<}$ of the transpose of matrix $\sigma$ should
satisfy, \be \label{27} \tilde{\nu}_{<} < \frac{1}{2}. \ee where
$\tilde{\nu}_{<}$ is defined as, \bea \label{28} \tilde{\nu}_{<}  =
\mathsf{min}[\tilde{\nu}_{+}, \tilde{\nu}_{-}~]; \eea and
$\tilde{\nu}_{\pm}$ is given by, \bea \tilde{\nu}_{\pm} =
\sqrt{\frac{\tilde{\Delta}(\sigma)\pm\sqrt{\tilde{\Delta}(\sigma)^{2}-4\mathsf{Det}\sigma}}{2}}
\quad; \eea where $\tilde{\Delta}(\sigma) = \Delta(\tilde{\sigma}) =
\mathsf{Det}(\alpha)+\mathsf{Det}(\beta)-2\mathsf{Det}(\mu)$. Thus
according to the condition (\ref{27}) when $\tilde{\nu}_{<} \ge1/2$
a Gaussian state become separable. The corresponding quantification
of entanglement is given by the logarithmic negativity
$E_{\mathcal{N}}$ \cite{Vidal,adeso, hor} defined as, \be \label{29}
E_{\mathcal{N}}(t) = \mathsf{max} [0, -\ln \{2\tilde{\nu}_{<}(t)\}];
\ee which constitute an upper bound to the \textit{distillable
entanglement} of any Gaussian state \cite{adeso}. On evaluating the
covariance matrix $\sigma$ for the state (\ref{20}) for $\gamma = 0$
(no loss), using equation (\ref{3}), (\ref{4}) and (\ref{21b}) we
find, \bea \label{30} \alpha = \beta =
\left[\begin{array}{cc} c & 0 \\
0 & d\end{array}\right];\qquad \mu =
\left[\begin{array}{cc} 0 & e \\
e & 0 \end{array}\right];
\eea
where $d, e, c$ are given by
\bea
\label{31}
c & = & \frac{1}{2}\{\cosh(2r)+\sinh(2r)\cos(2Jt)\} ;\nonumber\\
d & = & \frac{1}{2}\{\cosh(2r)-\sinh(2r)\cos(2Jt)\} ;\nonumber\\
e & = & -\frac{1}{2}\sinh(2r)\sin(2Jt). \eea The corresponding
symplectic eigenvalues $\tilde{\nu}_{\pm}$ are then given by \be
\label{32} \tilde{\nu}_{\pm} = \sqrt{cd}\pm e \ee One can clearly
see from equations (\ref{29}), (\ref{31}) and (\ref{32}) the
dependence of logarithmic negativity $E_\mathcal{N}$ on coupling
strength $J$ between the waveguides and the squeezing parameter $r$.
In figure 5. we plot the logarithmic negativity as a function of
scaled time, $\theta = Jt$ for the state $|\zeta\rangle$. Here $t$
is related to the length $l$ of the waveguide and its refractive
index $n$ by $t = nl/v$, $v$ being the velocity of light. We see
from figure (5) that as $|\zeta\rangle$ is separable at $t = 0$,
$E_\mathcal{N} = 0$ initially but as $Jt$ increases, it oscillates
periodically between a non-zero and zero value. Thus the initially
separable state $|\zeta\rangle$ becomes periodically entangled and
disentangled as its propagates through the waveguide. We attribute
this periodic generation of entanglement to the coupling $J$ among
the waveguides. We further find that $\tilde{\nu}_< = 1/2$ at
certain points along the waveguide given by $2\theta = (k+1)\pi, k =
0, 1,2,3,.....$. Note that at this points $E_\mathcal{N}$ vanishes
and $|\zeta\rangle$ becomes separable. At all other points the state
$|\zeta\rangle \neq |\zeta_{a}\rangle\otimes|\zeta_{b}\rangle$. We
see that $E_\mathcal{N}$ is maximum and has a value equal to the
amount of squeezing $2r$ at the points given by $2\theta =
(k+1)\pi/2$. Hence at this points the initial seperable state
$|\zeta\rangle$ becomes maximally entangled and is given by, \bea
\label{32a} |\zeta\rangle \equiv
\mathsf{exp}\{e^{i\pi}r(a^{\dagger}b^{\dagger}+ab)\}|00\rangle \eea
\begin{figure}
\begin{center}
\includegraphics[scale = 0.35]{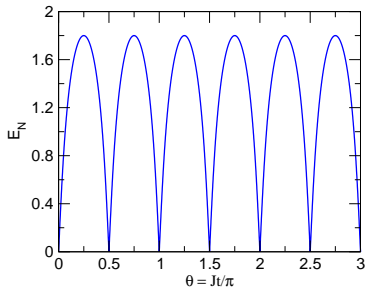}
\caption{ Plot of the time dependent logarithmic negativity
$E_\mathcal{N}$ for the state $|\zeta\rangle$. Here amount of
squeezing is taken to be $r = 0.9$.  }
\end{center}
\end{figure}

\subsection{Entangled two mode squeezed state as an input}

\noindent Let us now study the dynamical evolution of a two mode
squeezed state $|\xi\rangle$ as an input to the wave\-guide, \be
\label{33} |\xi\rangle =\mathsf{exp}[r(a^{\dagger}b^{\dagger}-
ab)]|00\rangle
 \ee
As before we consider $r$ to be real. To quantify the entanglement of the
state $|\xi\rangle$ we need to evaluate the logarithmic negativity
$E_{\mathcal{N}}$. Thus we first evaluate the covariance matrix
$\sigma$ for the state $|\xi\rangle$ using equations (\ref{3}) with $\gamma =
0$, (\ref{4}) and (\ref{21b}). We find $\sigma$ to be
\bea
\label{33a}
\sigma =
\left[\begin{array}{cccc} f & g & h & 0\\
g & f & 0 & -h\\
h & 0 & f & g\\
0 & -h & g & f\end{array}\right] \eea where $f,g$ and $h$ are given
by, \bea \label{34}
f & = & \frac{1}{2}\cosh(2r)\nonumber\\
g & = & -\frac{1}{2}\sinh(2r)\sin(2Jt)\nonumber\\
h & = & \frac{1}{2}\sinh(2r)\cos(2Jt); \eea

\begin{figure}[!h]
\begin{center}
\includegraphics[scale = 0.35]{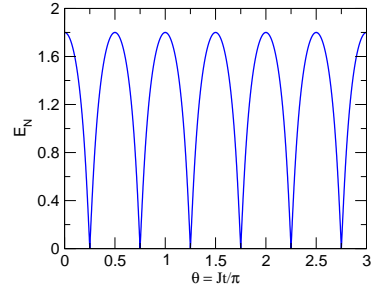}
\caption{Time evolution of logarithmic negativity $E_\mathcal{N}$
for the initial entangled state $|\xi\rangle$. Here the squeezing is
taken to be $r = 0.9$. We see that $E_\mathcal{N}$ is $\pi/4$ out of
phase to that for the state $|\zeta\rangle$. }
\end{center}
\end{figure}

The corresponding symplectic eigenvalues $\tilde{\nu}_{\pm}$ is then
given by, \bea \label{35} \tilde{\nu}_{\pm} &= &\sqrt{(f+g)(f-g)}\pm
h ; \eea The logarithmic negativity $E_{\mathcal{N}}$ can then be
evaluated using equations (\ref{28}), (\ref{29}) and (\ref{35}).
From equations (\ref{34}) and (\ref{35}) the dependence of
$E_{\mathcal{N}}$ on the squeezing $r$ and the coupling $J$ between
the wave\-guides is clearly visible . From equation (\ref{35}) we
find that $E_\mathcal{N} = 0$ \textit{i.e.} entanglement become zero
when, $2\theta = (k +1)\pi/2$ as then $\tilde{\nu}_{<} = 1/2$ and
thus the initially entangled state $|\xi\rangle$ becomes separable,
i.e $|\xi\rangle = \mathsf{exp}\{\frac{r}{2}e^{i\pi}( a^{\dagger
2}+a^{2})\}|0\rangle\otimes\mathsf{exp}\{\frac{r}{2}e^{i\pi}(b^{\dagger
2}+b^{2})\}|0\rangle$. In figure (6) we plot the time evolution of
$E_\mathcal{N}$ for $r = 0.9$. We see that entanglement oscillates
periodically between zero and non-zero values. We further find that
in this case the oscillations in $E_\mathcal{N}$ is $\pi/4$ out of
phase to that for the initial separable state $|\zeta\rangle$ . This
oscillatory behavior of entanglement is as discussed before, due to
the coupling $J$ among the wave\-guides.  Each time the states get
separable the presence of coupling leads to interaction among the
modes of the waveguides and creats back the entanglement. We see
from the figure that logarithmic negativity $E_\mathcal{N}$ reaches
maximum at later times at the points $2\theta = (k+1)\pi$ and is
equal to $2r$. Thus at this points the state $|\xi\rangle$ regains
its initial form given by equation (\ref{33}).

\section{Lossy Wave\-guides}

In this section we study the entanglement dynamics of lossy
wave\-guides ($\gamma\neq 0$). The loss $\gamma$ arises from the
loss in the material of the wave\-guide. In this case the dynamical
evolution of the wave\-guide modes is governed by the full
quantum-Louiville equation (\ref{3}). We next consider the cases of
both photon number state and squeezed states at the input of the
wave\-guide and discuss the influence of the loss on their
respective entanglement evolution.

\subsection{Effect of Loss on non-Gaussian Entanglement}

\noindent As discussed above, we first study the effect of loss on
the entanglement dynamics of the waveguide modes for photon number
input states. For this purpose we consider a single photon input
state $|1,1\rangle$ as the initial state. In this case we can
analytically solve the quantum-Louiville equation described in
(\ref{3}). To proceed further, we work in the interaction picture
such that the density matrix in the interaction picture is $
\tilde{\rho} (t) = e^{iJ t(a^\dagger
 b + b^\dagger
 a )} \rho (t)  e^{-iJ t(a^\dagger
 b + b^\dagger
 a )}$. In the interaction picture we
can write Eq. (\ref{3}) as
\begin{eqnarray}
\label{4444} \frac{\partial \tilde{\rho} (t) }{\partial t} & =
&-\frac{\gamma}{2} (\tilde{a}^{\dagger} \tilde{a}\tilde{\rho}-2
\tilde{a}
\tilde{\rho} \tilde {a}^{\dagger}+\tilde{\rho} \tilde{a}^{\dagger} \tilde{a})\nonumber\\
& & -\frac{\gamma}{2} (\tilde{b}^{\dagger}\tilde{b}\ \tilde{\rho}-2
\tilde{b} \tilde{\rho} \tilde{b}^{\dagger}+\tilde{\rho}
\tilde{b}^{\dagger} \tilde{b})~,
\end{eqnarray}
\noindent where $\tilde{a}$ and $\tilde{b}$ are given by
\begin{eqnarray}
\tilde{a}\hspace{0.01 cm}(t) & = & a \cos(Jt)-i b \sin(Jt)\nonumber\\
\tilde{b} \hspace{0.01 cm} (t) & = & b \cos(Jt)-ia \sin(Jt).
\end{eqnarray}

\noindent Using the above equation, we can rewrite Eq. (\ref{4444})
as
\begin{eqnarray}
\label{4445} \frac{\partial {\tilde{\rho}} (t)  }{\partial t} & = &-
\frac{\gamma}{2} (a^{\dagger} a \tilde{\rho} - 2 a \tilde{\rho}
a^{\dagger}+
\tilde{\rho} {a}^{\dagger} {a})\nonumber\\
& & -\frac{\gamma}{2} (b^{\dagger} b \tilde{\rho} - 2 b \tilde{\rho}
b^{\dagger}+ \tilde{\rho} {b}^{\dagger} {b}) ~.
\end{eqnarray}
\begin{widetext}

\noindent For the separable input state $|1,1\rangle$, the solution
for the density matrix (\ref{4445}) can be written as \cite{barnett}
:

\begin{eqnarray}
{\tilde{\rho}} (t) & = &   e^{ -4 \gamma t } \{(e^{2 \gamma
t}-1)^{2} |0,0\rangle \langle 0,0 |+(e^{2 \gamma t}-1) |1,0\rangle
\langle 1,0 | +(e^{2 \gamma t}-1) |0,1\rangle \langle 0,1 | +
|1,1\rangle \langle 1,1 |\}~.
\end{eqnarray}
\end{widetext}

\noindent Further, we can write ${\rho} (t) $ in terms
${\tilde{\rho}} (t) $ using the following equation:

\begin{eqnarray}
\rho (t) & = & e^{-iJ t(a^\dagger
 b + b^\dagger
 a ) }  \tilde{\rho}(t)  e^{i J t(a^\dagger
 b + b^\dagger
 a )}~.
\end{eqnarray}

\begin{figure}[htp]
 \scalebox{0.48}{\includegraphics{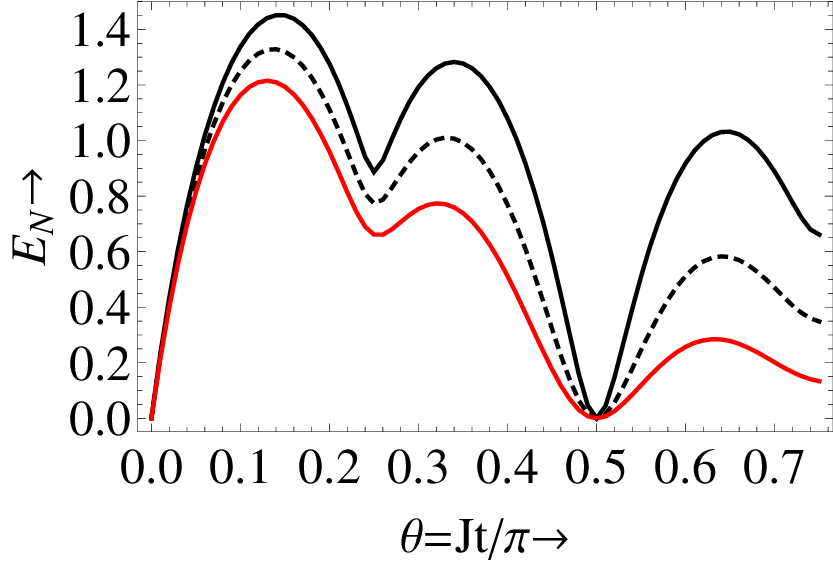}}
 \caption{\label{Fig3(a)} (Color online) Time evolution of the logarithmic negativity $E_\mathcal{N}$ in
presence of loss of the wave\-guide modes for the initial separable
input state (\ref{004}). The decay rates of the modes are given by
$\gamma/J = 0.1$ (solid black), $\gamma/J = 0.2$ (broken black) and
$\gamma/J = 0.3$ (red). }
\end{figure}

\noindent The above equation gives the time evolution of the density
matrix corresponding to the single photon state $|1,1\rangle$.
Following a similar approach as discussed in Sec. III, we can
evaluate the log negativity for the lossy wave\-guide case. But the
resulting expressions are lengthy and do not exhibit a simple
structure. Thus we only give the numerical results for the lossy
wave\-guide case. In Fig.~(\ref{Fig3(a)}), we show the decay of
entanglement, as a function of scaled time for the state
(\ref{004}). Note that the range of $\gamma/J$ values studied here
are similar to the numerical values used in the experiments
\cite{Stegeman,Aitchison}. For example, the coupling parameter $J$
for the lithium niobate waveguide lie between $1.83\times 10^{10}$
$sec^{-1}$ and $4.92\times 10^{10}$ $sec^{-1}$. The loss parameter
for these wave\-guides is close to $3 \times 10^9$ $sec^{-1}$
\cite{Stegeman} which corresponds to a value of $\gamma/J$ between
$1/7$ and $1/20$. For AlGaAs wave\-guides the loss $\gamma$ is close
to $2.7 \times 10^{10}$ $sec^{-1}$ \cite{Aitchison}. The coupling
parameter $J$ for these wave\-guides is about $2.46 \times 10^{11}$
$sec^{-1}$. Thus the $\gamma/J$ value for these wave\-guides is of
the order of $1/10$. It is worth mentioning that the $\gamma/J$
value for silica wave\-guides is significantly lower than the
corresponding values for the lithium niobate and AlGaAs
wave\-guides. This means that even a small loss would add up to a
significant decoherence in these complex quantum systems. From
Fig.~(\ref{Fig3(a)}) we find that for the lossy wave\-guide case the
entanglement between the wave\-guide modes decrease with time. In
addition, we find that increasing the value of $\gamma/J $ makes the
waveguide modes more fragile, as is evident from
Fig.~(\ref{Fig3(a)}). However, we find that the decrease in
entanglement is not substantial. Our results indicate that the
waveguide system can sustain the entanglement even for the higher
decay rates. Thus the coupled wave\-guide system can be used as an
efficient tool for the study of basic quantum optical effects. In
addition, the persistence of entanglement suggests that the coupled
wave\-guide system can be used effectively for various applications
in quantum information processing \cite{Matthews}. For example, the
single photon entanglement described here is a key step for the
successful implementation of the CNOT gate \cite{politi}. We also
studied the behavior of log negativity for the entangled initial
state (\ref{eq5}). In this case also we found that the entanglement
quantified by $E_{\mathcal{N}}$ shows a considerable robustness
against the decoherence effect.

\subsection{Effect of Loss on Gaussian Entanglement }

For the input squeezed state $|\zeta\rangle$ of equation (\ref{20})
we find that elements of the covariance matrix $\sigma$ in presence
of loss become dependent on the decay rate $\gamma$  and is given
by, \bea \label{37} \sigma =
\left[\begin{array}{cccc} c^{\prime} & 0 & 0 & e^{\prime}\\
0 & d^{\prime} & e^{\prime} & 0\\
0& e^{\prime}& c^{\prime}& 0\\
e^{\prime}& 0 & 0& d^{\prime}
\end{array}\right];\qquad
\eea where $c^{\prime}, d^{\prime}, e^{\prime}$ are given by \bea
\label{38}
c^{\prime} & = & \frac{1}{2}\{1+e^{-2\gamma t}\sinh^{2}(r)+e^{-2\gamma t}\sinh(2r)\cos(2Jt)\} ;\nonumber\\
d^{\prime} & = & \frac{1}{2}\{1+e^{-2\gamma t}\sinh^{2}(r)-e^{-2\gamma t}\sinh(2r)\cos(2Jt)\} ;\nonumber\\
e^{\prime} & = & -\frac{1}{2}e^{-2\gamma t}\sinh(2r)\sin(2Jt). \eea

The corresponding symplectic eigenvalue $\tilde{\nu}$ of the
covariance matrix is then found to be, \bea \label{39}
\tilde{\nu}_{\pm} = \sqrt{c^{\prime}d^{\prime}}\pm e^{\prime} \eea
\begin{figure}
\begin{center}
\includegraphics[scale = 0.45]{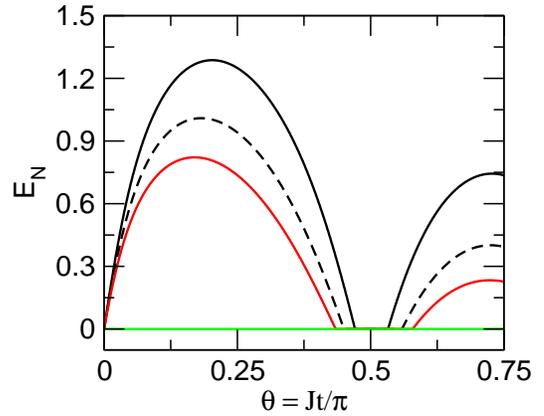}
\caption{(Color online) Time evolution of the logarithmic negativity
$E_\mathcal{N}$ in presence of loss of the wave\-guide modes for the
input state $|\zeta\rangle$. The decay rates of the modes are given
by $\gamma/J = 0.1$ (solid black), $\gamma/J = 0.2$ (broken black)
and $\gamma/J = 0.3$ (red). Here the squeezing is taken to be $r =
0.9$. The loss leads to new behavior in the entanglement.}
\end{center}
\end{figure}

On substituting equation (\ref{39}) in equations (\ref{28}) and
using (\ref{29}) we get the logarithmic negativity-for lossy
wave\-guides. To study the dependence of entanglement on loss of the
wave\-guide modes we plot the logarithmic negativity $E_\mathcal{N}$
for different decay rates $\gamma/J$ in figure (8). As for the case
of single photon states we focus on the range of $\theta$ important
from the experiment  point of view. We see new features in the
entanglement dynamics as an effect of the loss. We see from figure
(8) that in presence of loss the maximum value of entanglement for
the state $|\zeta\rangle$ reduces in comparison to the case of
loss\-less wave\-guides. However it is important to note that this
decrease is not substantial. We further find that with increase in
decay rate, the entanglement maximum shifts but does not show
considerable reduction (the maximum changes by only $0.4$ as the
decay rate becomes three times). Thus we see that entanglement is
quite robust against decoherence in this coupled wave\-guide
systems. The robustness of entanglement dynamics is an artifact of
coherent coupling among and the wave\-guide modes. This findings
hence suggest that coupled wave\-guide can be used as an effective
quantum circuit for use in  quantum information computations.
Further we see another new feature in entanglement in figure (8). We
find that there exist an interval of $\theta$ during which the state
$|\zeta\rangle$ remains separable. Note that in absence of loss the
state $|\zeta\rangle$ becomes separable momentarily and entanglement
starts to build up instantaneously once it becomes zero (see figure
5.) Thus this feature that entanglement
remains zero for certain interval of time arises solely due to loss. \\
In figure (9) we plot the long time behavior for entanglement of the
state $|\zeta\rangle$ with very small decay rate of $\gamma/J = 0.1$
and squeezing parameter $r = 0.9$. We see that entanglement decays
slowly with increasing $\theta$ as the magnitude of $E_\mathcal{N}$
diminish successively with every oscillations. In addition periods
of disentanglement arises repeatedly in its oscillations. We find
that the length of this periods increases with increasing $\theta$.
It is worth mentioning here that this kind of behavior has been
predicted earlier for two qubit entanglement \cite{das}.
\begin{figure}
\begin{center}
\includegraphics[scale= 0.45]{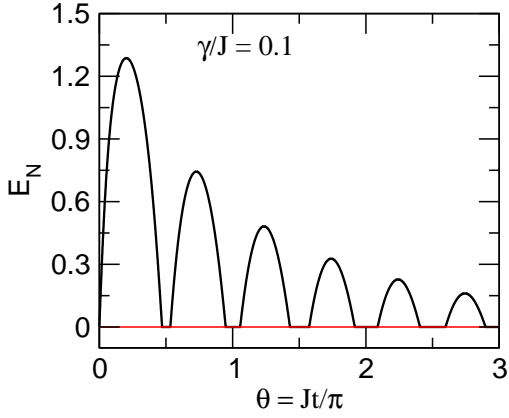}
\caption{Long time behavior of the logarithmic negativity
$E_\mathcal{N}$ in presence of loss of the wave\-guide modes for the
initial separable input state $|\zeta\rangle$.}
\end{center}
\end{figure}

Next we study the effect of the decay of wave\-guide mode on the
entanglement dynamics of the initial entangled squeezed state
$|\xi\rangle$ given in equation (\ref{33}). We find in this case the
covariance matrix to be,

\begin{figure}
\vspace{0.3 in}
\begin{center}
\includegraphics[scale= 0.45]{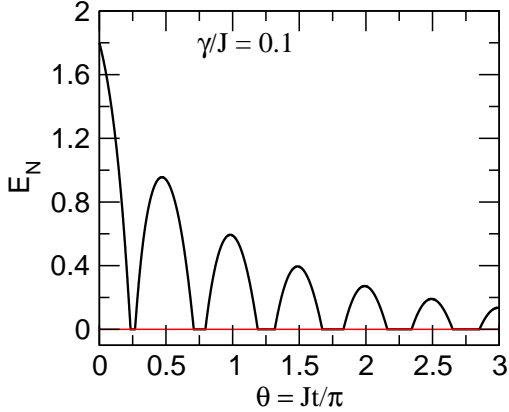}
\caption{Time evolution of the logarithmic negativity
$E_\mathcal{N}$ in presence of loss of the wave\-guide modes for the
initial entangled input state $|\xi\rangle$. Here $\gamma$ is the
decay rate of the modes and squeezing is taken to be $r = 0.9$.}
\end{center}
\end{figure}
\bea \label{40} \sigma =
\left[\begin{array}{cccc} f^{\prime} & g^{\prime} & h^{\prime} & 0\\
g^{\prime} & f^{\prime} & 0 & -h^{\prime}\\
h^{\prime} & 0 & f^{\prime} & g^{\prime}\\
0 & -h^{\prime} & g^{\prime} & f^{\prime} \end{array}\right] \eea
where $f^{\prime}, g^{\prime}, h^{\prime}$ are given by, \bea
\label{41}
f ^{\prime}& = & \frac{1}{2}+e^{-2\gamma t}\sinh^{2}(r)\nonumber\\
g^{\prime} & = & -\frac{1}{2}e^{-2\gamma t}\sinh(2r)\sin(2Jt)\nonumber\\
h^{\prime} & = & \frac{1}{2}e^{-2\gamma t}\sinh(2r)\cos(2Jt); \eea

In this case we find that the symplectic eigenvalues
$\tilde{\nu}_{\pm}$ are dependent on the decay rate of the
wave\-guide modes and is given by, \be \label{42} \tilde{\nu}_{\pm}
= \sqrt{m_{+}m_{-}}\pm h^{\prime} \ee where $m_{\pm}(t) =
1-e^{-2\gamma t}[1-\{\cosh(2r)\pm\sinh(2r)\sin(2Jt)\}] $. The
corresponding measure of entanglement given by the logarithmic
negativity $E_{\mathcal{N}}$ can then be calculated by using
equation (\ref{42}), (\ref{28}) and (\ref{29}). In figure (10) we
plot the logarithmic negativity $E_\mathcal{N}$ for the state
$|\xi\rangle$ as a function of $\theta$ in presence of loss. We find
similar behavior in the entanglement dynamics as seen earlier for
the separable state $|\zeta\rangle$. We find in figure (10) that
entanglement of the state $|\xi\rangle$ decrease slowly with
increasing $\theta$ for non-zero $\gamma/J$. Thus as for the
separable states, in case of initial entangled input states
entanglement is found to be quite robust in the face of loss. In
addition to this we also see in figure (10) periods of
disentanglement appearing successively as $\theta$ increases.

The loss in wave\-guides that we discussed in this section arises
due to material properties like change in refractive index and
absorption. On the other hand there can be decay of the waveguide
modes in the form of leakage to its surrounding also. It should be
noted that leakage is inherently different from the evanescent
coupling as the former can arise due to scattering and refraction
due to refractive index difference at the waveguide boundaries. Thus
the analysis of this section is also valid when the leakage is
important  as for example is the case when one couples channel
wave\-guides to slab wave\-guides \cite{Longhi3, Longhi44}.

\section{Conclusion}

\indent{} To conclude, we investigated the time evolution of
entanglement in a coupled wave\-guide system. We quantified the
degree of entanglement between the wave\-guide modes in terms of
logarithmic negativity. We have given explicit analytical results
for logarithmic negativity in case of initially separable single
photon states and for separable as well as entangled squeezed
states. We have also addressed the question of decoherence in
coupled wave\-guide systems by considering loss of wave\-guide
modes. For the lossy wave\-guides we found that the entanglement
shows considerable robustness even for substantial loss. Note that
our results are based on experimental parameters and thus should be
relevant for applications of wave\-guides in quantum information
sciences. Our results serve as guide for experiments dealing with
entanglement in wave\-guide structures. For efficient use of these
wave\-guides, one should choose the waveguide parameter like $\theta$ so
that one is away from values where the entanglement is minimum.\\

\vspace{1 cm}

\appendix
\section{TIME EVOLUTION OF THE INITIAL NOON STATE ($ |N,0\rangle+
|0,N\rangle/\sqrt{2}$)}

\begin{widetext}

\noindent In this appendix we give the details of our calculation
for $|\psi_ {out}\rangle $ when the input state is given by

\begin{eqnarray}
|\psi_ {in} \rangle  = \frac{( |N,0\rangle+
|0,N\rangle)}{\sqrt{2}}=\frac{( (a(0))^N + (b(0))^N )
|0,0\rangle)}{\sqrt{2 N!}}~,
\end{eqnarray}

\noindent Using Eq.~(\ref{4}) we can show that the input state given
by $ |\psi_ {in} \rangle$ evolves into a state~:

\begin{eqnarray}
|\psi_ {out}\rangle  = \frac{( (a(t))^N + (b(t))^N )
|0,0\rangle)}{\sqrt{2 N!}}~,
\end{eqnarray}

\noindent where $a(t)$ and $b(t)$ are given by Eq.~(\ref{4}). Using
Eq.~(\ref{4}) in the above equation, we get

\begin{eqnarray}\label{app}
&& |\psi_ {out}\rangle  = (\sum \beta_{k} |k,N-k\rangle)~,\nonumber\\
&&  \beta_{k} = \alpha_{k}+ \alpha_{N-k}~,\nonumber\\
&& \alpha_{k} = (C(N,k))^{1/2} (\cos(J t))^k (-i \sin(J t))^{N-k},
\end{eqnarray}

\noindent where $C(N,k)$ is the Binomial coefficient given by:
$C(N,k) = N!/(N-k)! k!$~.

\end{widetext}

\noindent This work was supported by NSF grant no CCF-0829860.

\end{document}